\documentclass[reprint, prl, aps, floatfix, superscriptaddress, twocolumn, showkeys]{revtex4-1}

\usepackage{amssymb}
\usepackage{amsmath}
\usepackage{graphicx}
\usepackage{bm}
\usepackage{dcolumn}
\usepackage{color}
\usepackage{ulem}
\usepackage{float}
\usepackage{textcomp}
\restylefloat{table}
\usepackage{verbatim}
\usepackage{epstopdf}
\usepackage[colorlinks]{hyperref}
\usepackage[mediumspace,mediumqspace,squaren]{SIunits}

\begin{document}

\title{Fractional quantum Hall valley ferromagnetism in the extreme quantum limit}
\date{\today}

\author{Md.\ Shafayat Hossain}
\author{Meng K.\ Ma}
\author{Y. J.\ Chung}
\author{S. K.\ Singh} 
\author{A. \ Gupta} 
\author{K. W.\ West}
\author{K. W.\ Baldwin}
\author{L. N. \ Pfeiffer}
\affiliation{Department of Electrical Engineering, Princeton University, Princeton, New Jersey 08544, USA}

\author{R.\ Winkler}
\affiliation{Department of Physics, Northern Illinois University, DeKalb, Illinois 60115, USA}

\author{M.\ Shayegan}
\affiliation{Department of Electrical Engineering, Princeton University, Princeton, New Jersey 08544, USA}


\begin{abstract}

Electrons’ multiple quantum degrees of freedom can lead to rich physics, including a competition between various exotic ground states, as well as novel applications such as spintronics and valleytronics. Here we report magneto-transport experiments demonstrating how the valley degree of freedom impacts the fractional quantum states (FQHSs), and the related magnetic-flux-electron composite fermions (CFs), at very high magnetic fields in the extreme quantum limit when only the lowest Landau level is occupied.  Unlike in other multivalley two-dimensional electron systems such as Si or monolayer graphene and transition-metal dichalcogenides, in our AlAs sample we can continuously tune the valley polarization via the application of \textit{in-situ} strain. We find that the FQHSs remain exceptionally strong even as they make valley polarization transitions, revealing a surprisingly robust ferromagnetism of the FQHSs and the underlying CFs. Our observation implies that the CFs are strongly interacting in our system. We are also able to obtain a phase diagram for the FQHS and CF valley polarization in the extreme quantum limit as we monitor transitions of the FHQSs with different valley polarizations. 

\end{abstract}

\maketitle 

The valley degree of freedom, i.e. the availability of multiple conduction-band minima,  opens new avenues for exciting physics and engineering. Similar to the spintronics, a new 'valleytronics' field has emerged, aiming to utilize the valley degree of freedom to make functional devices \cite{Gunawan.PRL.2006, Gunawan.PRB.2006, Shayegan.AlAs.Review.2006, Rycerz.2007, Zeng.NatNano.2012, Jones.NatNano.2013, Schaibley.2016, shafayat.valleybloch}. The role of valley degree of freedom in stabilizing different many-body ground states of low-disorder, two-dimensional electron systems (2DESs) is also of great fundamental importance \cite{Shayegan.AlAs.Review.2006, Shkolnikov.PRL.2002, Lai.2004, Shkolnikov.PRL.2005, Bishop.PRL.2007, Padmanabhan.PRB.2009, Padmanabhan.PRB.2010, Padmanabhan.PRL.2010, Gokmen.Natphy.2010, Young.NatPhy.2012, Feldman.2012, Feldman.2013, Shkolnikov.PRL.2002, Gokmen2.PRB.2010, Gokmen.ssc.2010}.  Of particular interest is quantum Hall valley ferromagnetism, namely the interaction-induced valley polarization of quantum Hall states, both at integer and fractional Landau level (LL) filling factors ($\nu$).  Here we study a fundamental characteristics of the valley degree of freedom, namely its persistence even in the extreme quantum limit where the 2D electrons occupy only the lowest LL ($\nu < 1$). We find that the fractional quantum Hall states (FQHSs) flanking $\nu=1/2$ are always present, implying a very robust quantum Hall valley ferromagnetism. As we tune the valley occupancy via applying \textit{in-situ} strain, the FQHSs make multiple transitions, reflecting the change in their valley polarization. 

\begin{figure*}[t!]
\includegraphics[width=.99\textwidth]{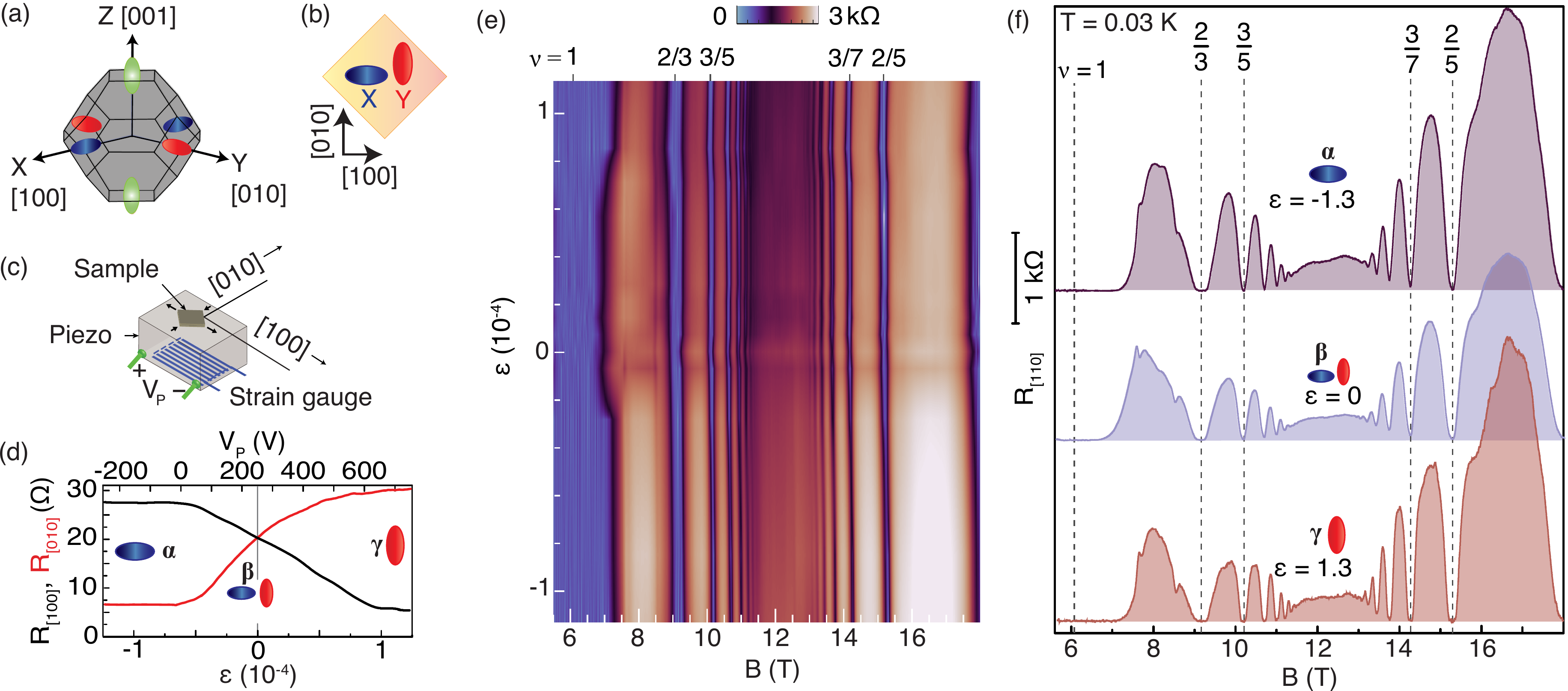}
\caption{\label{fig:Fig1} 
Sample description and characterization. (a) First Brillouin zone of bulk AlAs, showing anisotropic conduction band valleys. (b) Sample geometry, showing the orientation of the two occupied valleys (X and Y) and the measured resistances ($R_{[100]}$ and $R_{[010]}$). (c) Experimental setup for applying in-plane strain ($\varepsilon$). (d) Piezoresistance of the sample at $B = 0$ and $T \simeq 0.03$ K, measured as a function of $\varepsilon$. \boldsymbol{$\alpha$} - \boldsymbol{$\gamma$} mark the valley occupancies.
(e) Magnetotransport data at $T\simeq 0.03$ K, measured along [110],  as a function of $\varepsilon$ showing numerous odd-denominator FQHSs in the lowest LL. Some of the LL fillings are marked in the top axis. (f) Magnetoresistance traces ($R_{[110]}$) for different valley occupancies Traces are shown at three strain values as indicated (in units of $10^{-4}$), which correspond to the cases when the 2D electrons occupy only X valley, only Y valley, and both X and Y valley equally. Well-developed FQHSs, some of which are marked with dashed vertical lines, are seen in all traces.
} 
\end{figure*}

\begin{figure*}[t!]
\includegraphics[width=.99\textwidth]{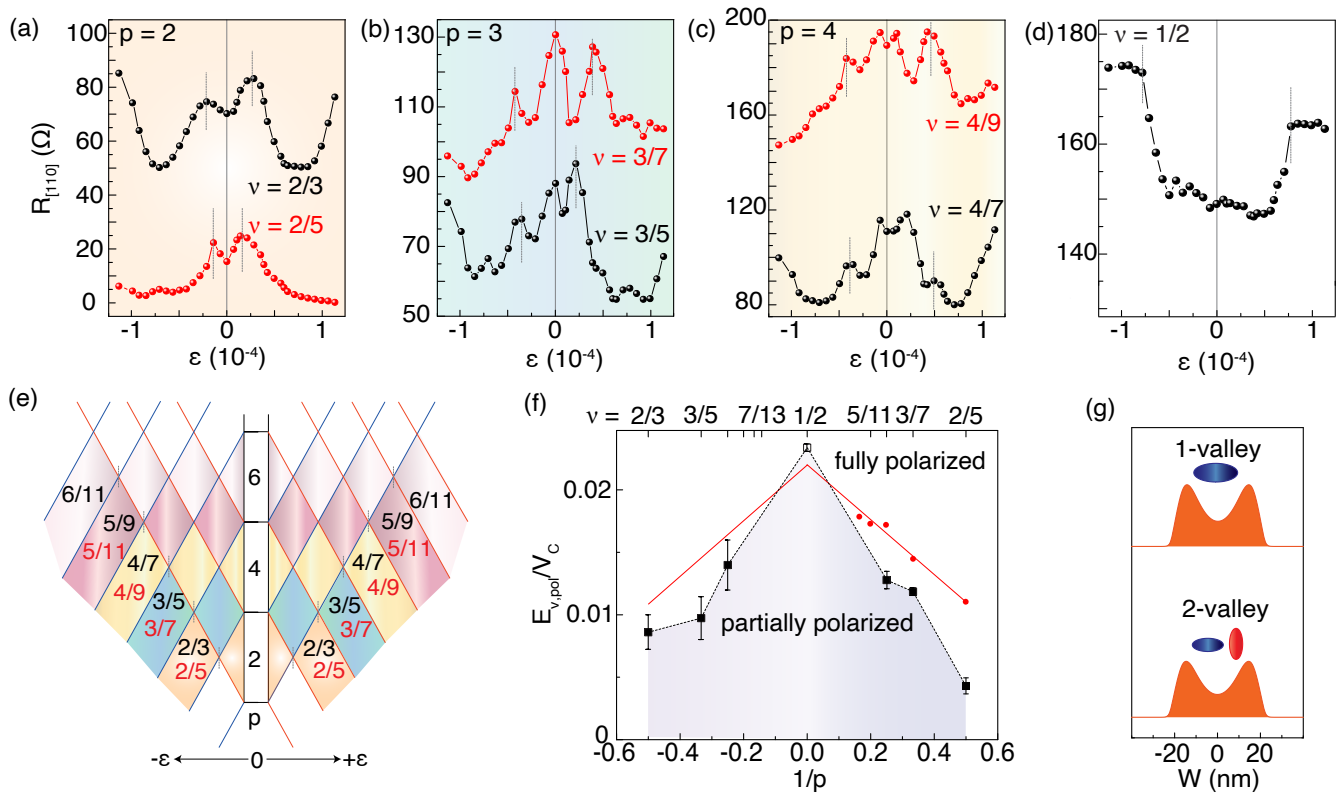}
\caption{\label{fig:Fig2} 
Valley transitions in the extreme quantum limit. (a-c) Piezoresistance along [110] at $p=2-4$, all taken at $T\simeq 0.3$ K, showing oscillations as a function of $\varepsilon$ until full valley polarization is reached. The $\varepsilon$ values for full valley polarization are marked with vertical dotted lines for different fillings. (d) Piezoresistances along [110] at $\nu=1/2$. (e) Lambda level ($\Lambda$L) diagram as a function of applied strain. $p$ denotes the $\Lambda$L index. The $\Lambda$Ls of the two valleys split as we apply $\varepsilon$ and instigate energy level coincidences and the consequent diamond-shaped regions as a function of $\varepsilon$; this in turn causes piezoresistance oscillations in panels (a-c). (f) Full valley polarization phase diagram in the extreme quantum limit extracted from the dotted vertical lines in panels (a-d).  These energies are normalized to the Coulomb energy for various FQHSs and the CF Fermi sea at $\nu=1/2$, and are plotted versus $1/p$.  The data points are obtained from the average $|\varepsilon|$ values for full valley polarization for $\varepsilon>0$ and $\varepsilon<0$, and the differences between the two sides are shown as the error bars. The data point at $\nu=1/2$ is based on $|\varepsilon|$ beyond which piezoresistance at $\nu=1/2$ saturates; see Fig. 2(d), also \cite{Footnote.1/2}. The black dotted lines connecting the data points represent the boundary between the partially and fully valley polarized phases.  Theoretically calculated spin polarization energy of CFs \cite{Park.PRL.1998} is shown using red circles (at positive $p$ values) and are joined by a red line.  A particle-hole symmetric red line is also drawn for negative $p$. (g) Charge distribution and Fermi sea for our AlAs sample when the electrons occupy single valley (top) and two valleys equally (bottom), obtained from self-consistent quantum mechanical calculations \cite{F1}.
} 
\end{figure*}

Our material platform is a 2DES, with density $n = 1.45 \times 10^{11}$ cm$^{-2}$, and mobility $\mu = 7.5 \times 10^{5}$ cm$^{2}/$Vs confined to a 45-nm-wide AlAs quantum well (QW). Electrons in bulk AlAs occupy three ellipsoidal valleys (X, Y, and Z) centered at the X point of the Brillouin zone with their major axes lying in the [100], [010], and [001] crystallographic directions, respectively [Fig. 1(a)]. However, this three-fold degeneracy is lifted when we grow an AlAs QW on a GaAs substrate because the biaxial compression originating from the lattice mismatch between AlAs and GaAs pushes the Z valley higher in energy relative to X and Y valleys (we denote the growth direction as [001]). For a sufficiently wide AlAs QW such as ours, the 2D electrons occupy only the X and Y valleys, with their major axes lying in the plane [Fig. 1(b)] \cite{Shayegan.AlAs.Review.2006, Chung.PRM.2018}. The 2D electrons in each valley possess an anisotropic Fermi sea with longitudinal and transverse effective masses of $m_l = 1.1m_0$ and $m_t = 0.20m_0$, leading to an in-plane effective mass of $ m^*= (m_t m_l)^{1/2}=0.46m_0$, where $m_0$ is the free electron mass. In the absence of any external strain that breaks the in-plane symmetry, the X and Y valleys are degenerate in energy. We can control the relative occupancy of X and Y by applying an in-plane, uniaxial strain $ \varepsilon=\varepsilon_{[100]}-\varepsilon_{[010]}$, where $\varepsilon_{[100]}$ and $\varepsilon_{[010]}$ are the strain values along [100] and [010] \cite{Shayegan.AlAs.Review.2006}. The valley-splitting energy is given by $E_V=\varepsilon E_2$, where $E_2$ is the deformation potential, which in AlAs has a band value of $5.8$ eV \cite{Shayegan.AlAs.Review.2006}. In our experiments, we glue the sample to a piezo-actuator, as displayed in Fig. 1(c), and apply a voltage bias ($V_P$) to its leads to control the amount of strain ($3.3 \times 10^{-7}$ V$^{-1}$). We used standard lock-in techniques for resistance measurements which were carried out in a dilution refrigerator and in a $^3$He cryostat with base temperatures of $T\simeq0.03$ K and $0.3$ K, respectively. 

Figure 1(d) demonstrates how we tune and monitor the valley occupancy. Here we show the sample's piezoresistance as a function of $V_{P}$. At point \boldsymbol{$\beta$}, the 2DES exhibits isotropic transport, namely, the resistances measured along [100] and [010] ($R_{[100]}$ and $R_{[010]}$) are equal, even though the individual valleys are anisotropic.  At this point $\varepsilon=0$ and the two valleys are degenerate. Note that a finite $V_P$ is often required to attain $\varepsilon=0$ because of a sample- and cooldown-dependent residual strain \cite{Shayegan.AlAs.Review.2006}. For $\varepsilon>0$, as electrons transfer from X to Y, $R_{[100]}$ decreases [black trace in Fig. 1(d)] because the electrons in Y have a small effective mass and therefore higher mobility along [100]. (Note that the total 2DES density remains fixed as strain is applied.) The resistance eventually saturates (point \boldsymbol{$\gamma$}), when all electrons are in Y \cite{Shayegan.AlAs.Review.2006, Gokmen.Natphy.2010, ShafayatAlAs.PRL.2018}. For $\varepsilon<0$, $R_{[100]}$ increases and saturates (point \boldsymbol{$\alpha$}) as the electrons are transferred to X which has a large mass and low mobility along [100]. As expected, the behavior of $R_{[010]}$ is the opposite of $R_{[100]}$; see the red trace in Fig. 1(d).  Such a continuous valley tuning is not possible in other multivalley systems, such as Si \cite{Lai.2004} or single-layer graphene \cite{Feldman.2012, Feldman.2013}.


When we apply a perpendicular magnetic field ($B$) to the sample, the kinetic energy of the 2DES is quenched and the density of states becomes quantized into a set of LLs. Our focus here is the extreme quantum limit where the lowest-energy LL is partially occupied ($\nu<1$). The physics of the 2DES is dominated by electron-electron interaction, as manifested by series of FQHSs at odd denominator fillings flanking $\nu=1/2$ \cite{Jain.2007, Halperin.Jain.2020}. We show in Fig. 1(e) magnetoresistance of the sample for $\nu<1$ as a function of $\varepsilon$ and $B$, covering the complete spectrum from only X ($\varepsilon=-1\times 10^{-4}$) to only Y ($\varepsilon=1 \times 10^{-4}$) valley occupancies. (Note that the resistance shown here, $R_{[110]}$, is measured along the [110] direction to minimize the impact of the change in mobility with valley occupancy on resistance \cite{Padmanabhan.PRB.2009}.) In Fig. 1(f) we show examples of magnetoresistance traces for the three cases marked in Fig. 1(d); \boldsymbol{$\alpha$}: only X, \boldsymbol{$\gamma$}: only Y, and \boldsymbol{$\beta$}: equal X and Y valley occupancies. Pronounced resistance minima, signaling well-developed FQHSs at $\nu=2/3, 3/5, 4/7, …$, and $1/3, 2/5, 3/7, …$ are seen on the two sides of $\nu=1/2$ throughout the applied strain range. The presence of such a rich set of FQHSs, up to $\nu=7/15$ and $7/13$, attests to the extremely high quality of our sample. 

An understanding of FQHS data can be provided by the composite fermion (CF) theory in which, at $\nu=1/2$, two magnetic flux quanta are attached to each electron to form a CF \cite{Jain.2007, Halperin.Jain.2020, Jain.PRL.1989, Halperin.Lee.Read.PRB.1993}. In a mean-field theory, the flux attachment cancels the applied magnetic field and, at $\nu=1/2$, the CFs form a Fermi sea. The FQHSs on the flanks of $\nu=1/2$ can then be explained as the integer quantum Hall states (IQHSs) of CFs. The key concept is that, akin to the LLs of electrons, CFs also form quantized energy levels, termed $\Lambda$-levels ($\Lambda$Ls) \cite{Jain.2007}, in the presence of an \textit{effective} magnetic field measured relative to the field at $\nu=1/2$. This maps a FQHS at electron LL filling $\nu$ to a CF $\Lambda$L filling $p$ according to $\nu=p/(2p+1)$. Now, if CFs have a valley degree of freedom, the FQHS near $\nu=1/2$ should follow the $\Lambda$L fan diagram schematically shown in Fig. 2(e) where the $\Lambda$Ls of the X valley are drawn in blue and those of the Y valley in red \cite{FootenoteB}. According to this diagram, FQHSs at different fillings should go through multiple transitions. For example, the FQHSs at $\nu=2/3$ and $2/5$ ($p=2$) should be strong at $\varepsilon=0$, become weak or disappear at a specific value of $|\varepsilon|$ where the X and Y valley $\Lambda$Ls cross, and then become strong again and remain strong at larger values of $|\varepsilon|$. In contrast, the FQHSs at $\nu=3/5$ and $3/7$ ($p=3$), should be weak at $\varepsilon=0$ and then, as a function of $|\varepsilon|$, become strong, weak, and then strong again. Such an evolution of FQHSs has been reported near $\nu=3/2$ \cite{Bishop.PRL.2007, Padmanabhan.PRB.2009}. A qualitatively similar \textit{electron} LL diagram has also been used to explain the evolution of the \textit{IQHSs} with strain \cite{Gunawan.PRL.2006}.
 
In our data taken at $T = 0.03$ K in the extreme quantum limit, at first sight it appears that the CFs have lost their valley degree of freedom, as the FQHSs remain quite robust and do not show any sign of transitions in Figs. 1(e) or (f) in the entire range of $\varepsilon$. However,  re-measuring the magnetoresistance traces at $T=0.30$ K, and carefully analyzing the evolution of the values of the FQHS resistance minima, reveal that the FQHSs do indeed undergo transitions. The higher temperature of 0.3 K lifts the resistances at FQHS fillings from zero and allows us to observe changes as a function of applied $\varepsilon$. The data are presented in Figs. 2(a-c) for $p=2$ ($\nu=2/3$ and $2/5$), $p=3$ ($\nu=3/5$ and $3/7$), and $p=4$ ($\nu=4/7$ and $4/9$). In these plots, the resistance at a given FQHS filling oscillates as a function of $\varepsilon$ and the number of oscillations depends on $\nu$. For example, at $\nu=3/7$ [Fig. 2(b)], the resistance is high at $\varepsilon=0$, reflecting a weak FQHS. As $|\varepsilon|$ is increased, the resistance first drops, signaling a stronger FQHS, then shows a maximum reflecting a weakening of the FQHS, and then eventually becomes small and remains small consistent with a strong and fully valley-polarized FQHS. This is exactly the evolution expected from Fig. 2(e) $\Lambda$L diagram \cite{FN}. Data at other fillings [Figs. 2(a,c)] also exhibit oscillations which are consistent with Fig. 2(e).

The results shown in Figs. 2(a-c) indicate that the FQHSs undergo transitions as their valley degree of freedom is tuned.  However, as seen in Figs. 1(e-f), there are always FQHSs at these fillings as manifested by the very strong magnetoresistance minima. (Strong magnetoresistance minima are also observed at 0.3 K \cite{SM}). This indicates that quantum Hall valley ferromagnetism, evinced by the presence of IQHS minima at LL crossings in \textit{interacting} 2DESs \cite{Shayegan.AlAs.Review.2006, Sondhi.PRB.1993, Shkolnikov.PRL.2005, Gokmen2.PRB.2010}, persists in the FQHS regime \cite{Padmanabhan.PRL.2010}, suggesting that there is interaction between CFs near $\nu=1/2$. It is worth mentioning that, near $\nu=3/2$, the FQHSs in AlAs 2DESs completely disappear or become very weak at $\Lambda$L crossings \cite{Padmanabhan.PRB.2009, Padmanabhan.PRL.2010}. This is also true in a high-mobility Si 2DES, where two out-of-plane valleys are occupied \cite{Lai.2004}. However, in contrast to the AlAs data near $\nu=1/2$ reported here, in the two-valley Si 2DES only the $\nu=1/3$ is observed and the other odd-numerator FQHSs (e.g., at $\nu=3/5$ and 3/7) are extremely weak or absent \cite{Lai.2004}. It is possible that the extremely high quality of the AlAs 2DESs is responsible for the very strong FQHSs we observe even at $\Lambda$L crossings and the implied robust FQH ferromagnetism.  Another difference is that the 2DES in Si occupies two out-of-plane (Z) valleys which have circular and overlapping Fermi seas once projected on the 2D plane, while in AlAs the electrons are in X and Y valleys which are anisotropic and have non-overlapping projected Fermi seas [Fig. 1(b)].

Note in Fig. 2(e) that, beyond the last $\Lambda$L coincidence, the system should be fully valley polarized. We can therefore determine the energy for full valley polarization ($E_{V,pol}$) of a given FQHS from the value of strain ($\varepsilon_{pol}$) for the last resistance maximum in Figs. 2(a-c) and using the expression $E_{V,pol} = E_2 \varepsilon_{pol}$, where $E_2 =5.8$ eV is the deformation potential for the AlAs conduction band \cite{Shayegan.AlAs.Review.2006, Padmanabhan.PRB.2009}. In Fig. 2(f), we summarize the measured valley polarization energies, normalized by the Coulomb energy $V_C=e^2/4\pi\kappa \varepsilon_0 l_B$, for different FQHSs; $\kappa$ is the dielectric constant of AlAs ($\simeq 10$) and $l_B=(\hbar/eB)^{1/2}$ is the magnetic length. In Fig. 2(f) we have also included a point for full valley polarization of CFs at $\nu=1/2$ from the value of $|\varepsilon|$ above which the piezoresistance at this filling saturates [Fig. 2(d)]. Qualitatively similar piezoresistance traces have been reported at $\nu=3/2$ and $1/2$ \cite{Padmanabhan.PRB.2009, Padmanabhan.PRB.2010, Gokmen.ssc.2010} and are attributed, partly, to the loss of screening as the CFs become valley polarized; see, e.g. \cite{Gokmen.ssc.2010, Footnote.1/2}. 

As seen in Fig. 2(f), the valley polarization energy increases for higher-order FQHSs as $\nu=1/2$ is approached. This is qualitatively similar to what has been reported for the valley polarization of FQHSs near $\nu=3/2$ in AlAs 2DESs \cite{Padmanabhan.PRB.2009}, or spin polarization in GaAs 2DESs \cite{Liu.PRB.2014}. It is also qualitatively consistent with what is theoretically expected for the \textit{spin} polarization of FQHSs and the underlying CFs \cite{Park.PRL.1998}; see the red lines in Fig. 2(f). (No theoretical calculations are available for the valley polarization of FQHSs.)

Despite the above qualitative similarities, there are two noteworthy features. First, in Fig. 2(f), except for the data point at $\nu=1/2$ \cite{Footnote.1/2}, the experimental data for the CF spin polarization are well below the theoretical boundary (red curves). This is qualitatively similar to the case of spin polarization of FQHSs in GaAs \cite{Liu.PRB.2014}, where the difference is partly attributed to the non-zero thickness of the electron layer and the finite energy separation between the LLs, namely, LL mixing \cite{Zhang.PRL.2016}. Note that the 2DES in our 45-nm-wide AlAs QW has a reasonably large layer thickness and the charge distribution is bilayer like $\big($see Fig. 2(g)$\big)$ \cite{F1}.  Moreover, we expect the large effective mass in AlAs to lead to significant LL mixing \cite{Footnote.LL.mixing}. The observed lower experimental boundary for full valley polarization compared to the calculated boundary in Fig. 2(f) is therefore not unexpected. Second, in our AlAs 2DESs, including the present sample, we find that the valley polarization energies for FQHSs near $\nu=3/2$ (e.g., at $\nu=4/3$) are about 2 to 3 times larger than the equivalent FQHSs near $\nu=1/2$ (e.g., at $\nu=2/3$) \cite{Padmanabhan.PRB.2010}. Similarly, the spin polarization energies are found to be larger near $\nu=3/2$ compared to $1/2$ in high-mobility GaAs 2DESs \cite{Liu.PRB.2014}. In both valley and spin cases, the disparity has been attributed to a breakdown of particle-hole symmetry, triggered by LL mixing \cite{Padmanabhan.PRB.2010, Liu.PRB.2014}.

Before closing we point out another noteworthy feature of our data. A recent theory \cite{Zhu.PRB.2018} predicts that the CFs at and near $\nu=1/2$ in 2DESs with a highly anisotropic effective mass should spontaneously valley polarize. This is analogous to the Stoner model of ferromagnetism \cite{Stoner.1947} except that here the transition is induced by anisotropy instead of density. Ref. \cite{Zhu.PRB.2018} suggests that the threshold of the effective mass anisotropy for the onset of spontaneous CF valley polarization is $\sim 7$. This is larger than the anisotropy in our 2DES where $m_l/m_t \simeq 5$. Nevertheless, Ref. \cite{Zhu.PRB.2018} concluded that the spontaneous valley polarization was indeed observed in AlAs 2DESs, based on the  $\nu=3/5$ experimental data of Ref. \cite{Padmanabhan.PRB.2010}. We emphasize that, our data shown in Fig. 2(b), which were taken on a much higher quality AlAs 2DES, clearly indicate that the 3/5 FQHS is \textit{not} fully valley polarized and requires a finite amount of strain for valley polarization.  Besides quality, however,  one should keep in mind that our 2DES here has a much larger layer thickness compared to the samples used in Ref.  \cite{Padmanabhan.PRB.2010} or considered in Ref.  \cite{Zhu.PRB.2018}.

\begin{acknowledgments}
We acknowledge support through the U.S. Department of Energy Basic Energy Science (Grant No. DEFG02-00-ER45841) for measurements, and the National Science Foundation (Grants No.  DMR 2104771 and No. ECCS 1906253), and the Gordon and Betty Moore Foundation’s EPiQS Initiative (Grant No. GBMF9615 to L. N. P.) for sample fabrication and characterization.  We also acknowledge QuantEmX travel grants from the Institute for Complex Adaptive Matter and the Gordon and Betty Moore Foundation through Grant No. GBMF5305 to M.S.H., M.K.M., and M.S. A portion of this work was performed at the National High Magnetic Field Laboratory, which is supported by the National Science Foundation Cooperative Agreement No. DMR-1644779 and the State of Florida. We thank S. Hannahs, T. Murphy, J. Park, H. Baek, and G. Jones at NHMFL for technical support. We also thank J. K. Jain for illuminating discussions.
\end{acknowledgments}

\pagebreak[4]
\clearpage
\begin{widetext}
\begin{center}
\textbf{\large Supplemental Material: Fractional quantum Hall valley ferromagnetism in the extreme quantum limit}
\end{center}
\end{widetext}
\setcounter{equation}{0}
\setcounter{figure}{0}
\setcounter{table}{0}
\setcounter{page}{1}
\makeatletter
\renewcommand{\theequation}{S\arabic{equation}}
\renewcommand{\thefigure}{S\arabic{figure}}
\renewcommand{\bibnumfmt}[1]{[S#1]}
\renewcommand{\citenumfont}[1]{S#1}

\section*{I. Methods}
\begin{figure}[b!]
\includegraphics[width=.495\textwidth]{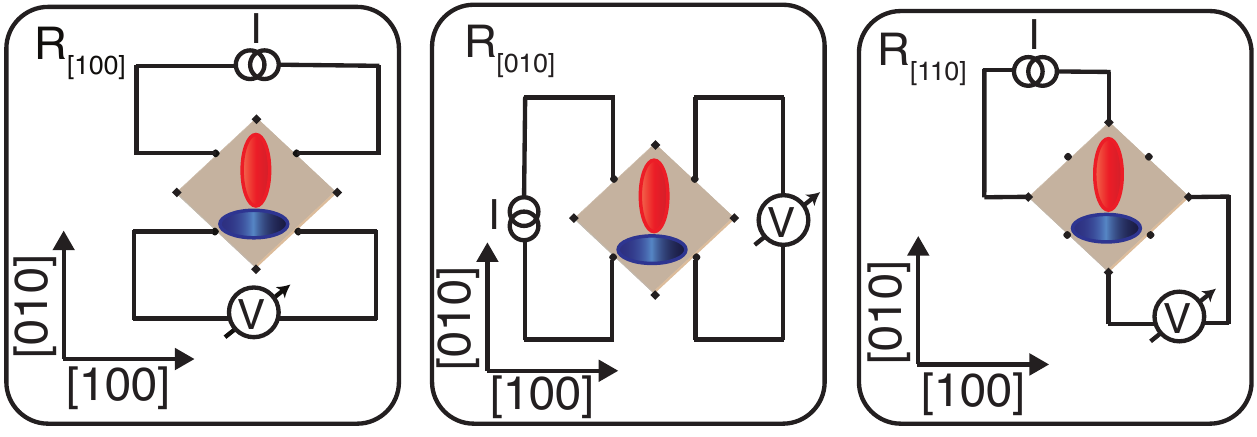}
\caption{\label{fig:FigS3} 
Configurations showing the current injection and voltage probes used to measure $R_{[100]}$, $R_{[010]}$, and $R_{[110]}$.}
\end{figure} 

In our measurements of piezoresistances along different crystallographic directions, we used very small van der Pauw samples with dimensions of $1.5$ mm $\times 1.5$ mm. The small size helps to maintain sufficient strain homogeneity. Note that the sample edges are along the GaAs cleave directions, $[110]$ and $[\bar110]$. The samples are lapped and polished on the backside down to $\simeq 150$ $\mu$m in order for the strain to propagate to the 2DES \cite{Shayegan.APL.2003}. Electron-beam-evaporated Ti-Au alloy on the backside of the sample shields the 2DES from the electric field generated by the applied voltage bias to the piezoelectric actuator on which the sample is glued.  We note that during the cooldown, due to the difference in the thermal contraction coefficient between the sample and piezo, a residual strain develops. We can compensate for this residual strain by applying an opposite uniaxial strain via the piezo actuator to attain the $\varepsilon=0$ condition. 

In Fig. S1 we show the sample voltage and current leads for all the configurations used in our magnetotransport measurements. We carried out our experiments partly in a dilution refrigerator with a mixing chamber temperature of $T\simeq0.03$ K and at the National High Magnetic Field Laboratory (NHMFL) in Tallahassee, Florida.  We also used a $^3$He system with a base temperature of 0.3 K.

To obtain the resistance values as a function of strain at different filling factors [plots of Figs. 2(a-d) of the main text], we change the strain at zero magnetic field and then sweep the field to obtain the magnetoresistance trace for a fixed strain. Thus, the discrete data points are obtained from the magnetoressiatnce traces at different strains.  This strategy is required for strain-dependent studies in our samples because our low density samples become inhomogeneous when they experience changes in strain at high magnetic fields.

We measured the resistance along [110] for our study. This is because [100] and [010] directions show a change in resistance by nearly a factor of 3 when electrons are transferred fully  from one valley to the other [Fig. 1(f) of the main text].Such a large change in resistance can mask the small features relevant to the Lambda level ($\Lambda$L) crossings. On the other hand, resistance along the [110] direction is most suitable for tracking the crossings because, in principle, there is no change in resistance along [110] as a function of valley occupancy as the mobility changes along [100] and [010] do not influence the mobility along [110]. Hence the small features relevant to the $\Lambda$L crossings can be resolved clearly; also see Refs.  \cite{Bishop.PRL.2007, Padmanabhan.PRB.2009, Padmanabhan.PRB.2010}.

\section*{II. Representative magneto-resistance traces at T = 0.3 K}

\begin{figure}[b!]
\includegraphics[width=.43\textwidth]{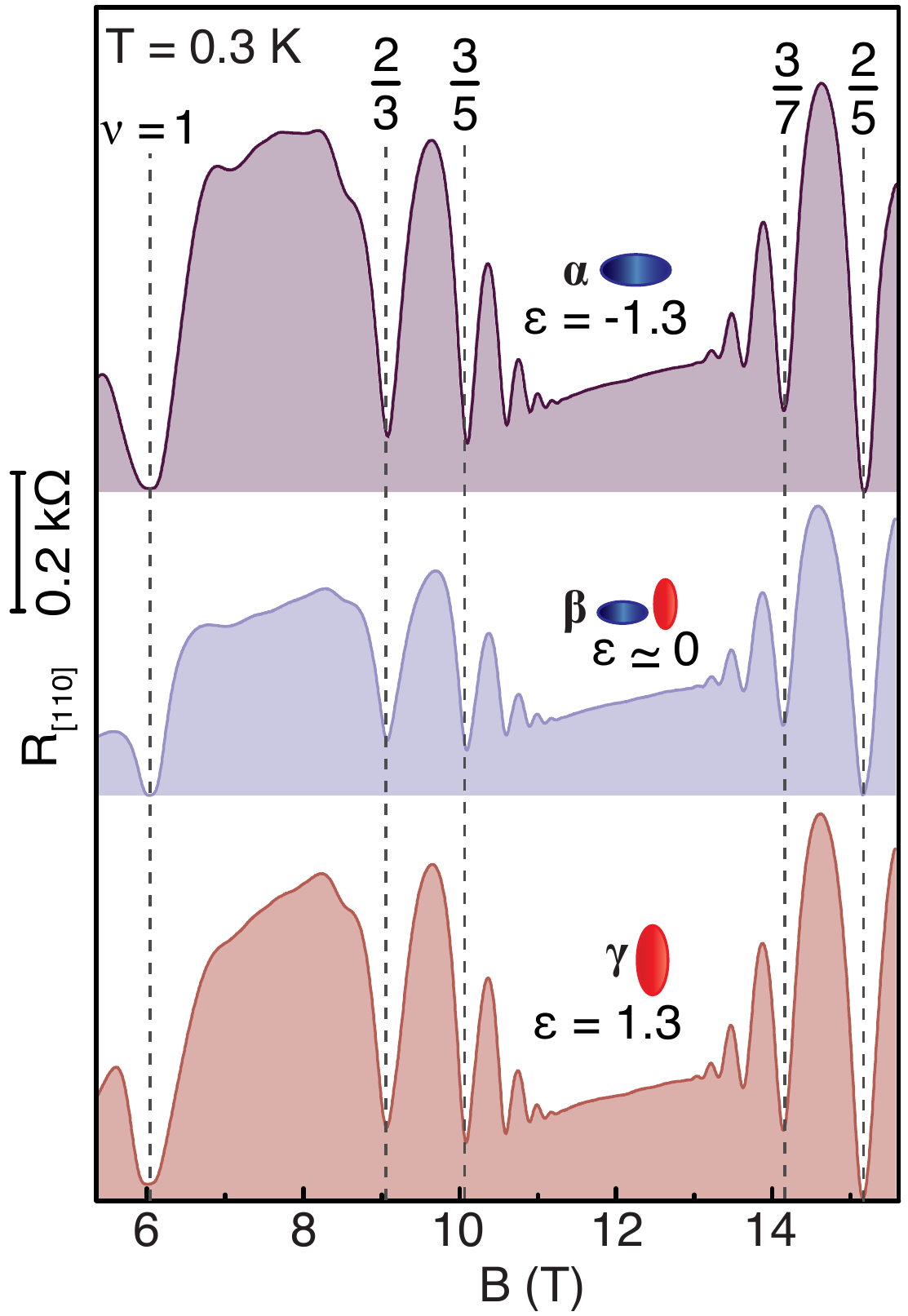}
\caption{\label{fig:Fig1} 
Magnetoresistance traces ($R_{[110]}$) taken at $T = 0.3$ K for different valley occupancies. Traces are shown at three strain values as indicated (in units of $10^{-4}$), which correspond to the cases when the 2D electrons occupy only the X valley, only the Y valley, and both X and Y valleys equally. Well-developed FQHSs signaled by resistance minima as marked, are seen for all three cases.
} 
\end{figure}

Figure S2 shows the magnetoresistance traces along the [110] crystallographic direction taken at $T=0.3$ K.  At $T=0.3$ K, the fractional quantum Hall states (FQHSs) up to $5/11$ and $7/13$, are seen. Such a rich set of FQHSs even at $T=0.3$ K attests to the extremely high quality of our sample. 

\section*{III. Valley transition in the fractional quantum Hall states as a function of strain}
\begin{figure}[t!]
\includegraphics[width=.35\textwidth]{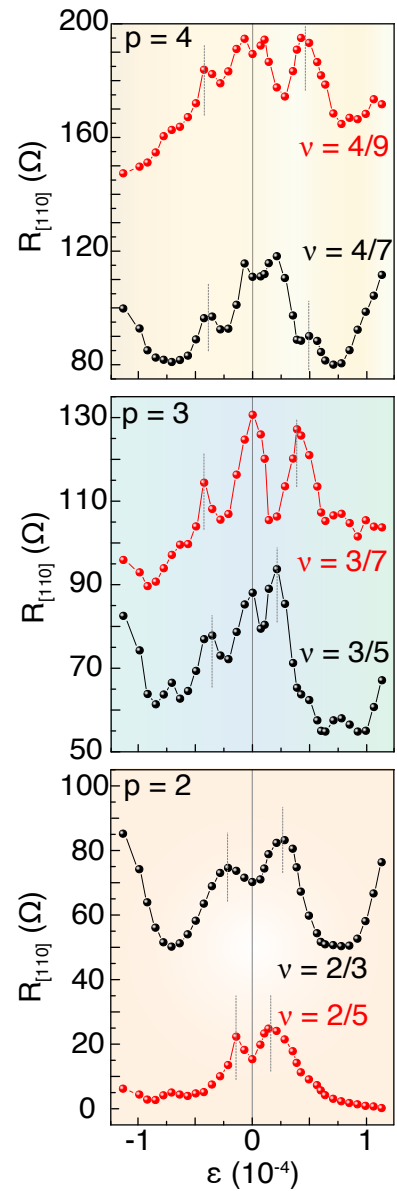}
\caption{\label{fig:Fig2} 
Piezoresistance along [110] at $p=2-4$, all taken at $T\simeq 0.3$ K. This is similar to Figs. 2(a-c) of the main text, except that now the traces are plotted in the same column, showing the relative positions of the oscillations as a function of $\varepsilon$ at different $p$. The $\varepsilon$ values for full valley polarization are marked with vertical dotted lines for different fillings. 
} 
\end{figure}

In this section, we first show our experimental data demonstrating the valley transitions in different FQHSs and then delineate how they fit into the picture of FQH ferromagnetism.

First, in Fig. S3, we show the piezoresistance plots at different $p$, also shown in Figs. 2(a-c) of the main text, now all in the same column to highlight relative positions of the valley transitions at different $p$. In order to pinpoint the location of the full valley polarization, we identified the last peak positions corresponding to the full valley polarization on the positive and negative strain sides of the data in Figs. 2(a-c). We used the average of the positions on the two sides. The error bars shown in Fig. 2(f) of the main text reflect the difference in the strain positions between the positive and negative sides. 

Next, we comment on the interpretation of the composite fermion (CF) and FQH valley transitions within the robust quantum Hall valley ferromagnetism that we observe even at the zero strain condition.  The valley transitions manifest as changes in resistance (at a fixed FQH filling and fixed $T=0.3$ K), which implies a change in the FQH energy gap. Here, the FQH energy gaps do not close; they only get stronger or weaker as the system make transitions between different valley polarizations. This is similar to and consistent with the case of spin transitions of CFs \cite{eisenstein.prb.1990, engel.prb.1992}. 

It is worthwhile noting that FQH valley ferromagnetism involves interaction between CFs and cannot be explained using a non-interacting CF picture or a simple Lambda level diagram as shown in Fig. 2(e) of the main text.  We add that quantum Hall ferromagnetism at $\nu=1$, i.e., the presence of a quantum Hall state in the case of spin \cite{Leadley.PRL.1997} or valley degeneracy \cite{Shkolnikov.PRL.2005} is a well-studied phenomenon.  A natural question is what happens to the FQHSs when there is quantum Hall ferromagnetism at $\nu=1$. In the case of spin,  FQHSs in the lowest Landau level still exhibit spin transitions \cite{eisenstein.prb.1990}. Here we observe a qualitatively similar phenomenon as the FQHSs undergo valley transitions even though a quantum Hall state at $\nu=1$ is present at all strains.


\begin{thebibliography}{99}

\bibitem{Gunawan.PRL.2006} O. Gunawan, Y. P. Shkolnikov, K. Vakili, T. Gokmen, E. P. De Poortere, and M. Shayegan, Valley Susceptibility of an Interacting Two-Dimensional Electron System, Phys. Rev. Lett. \textbf{97}, 186404 (2006).

\bibitem{Gunawan.PRB.2006} O. Gunawan, B. Habib,  E. P. De Poortere,  and M. Shayegan, Quantized conductance in an AlAs two-dimensional electron system quantum point contact, Phys. Rev. B \textbf{74}, 155436 (2006).

\bibitem{Shayegan.AlAs.Review.2006} M. Shayegan, E. P. De Poortere, O. Gunawan, Y. P. Shkolnikov, E. Tutuc, and K. Vakili,  Two-dimensional electrons occupying multiple valleys in AlAs, Phys. Stat. Sol. (b) \textbf{243}, 3629 (2006).

\bibitem{Rycerz.2007} A. Rycerz, I. Tworzydlo, and C. W. Beenakker, Valley filter and valley valve in graphene, Nature Physics \textbf{3}, 172 (2007).

\bibitem{Zeng.NatNano.2012} H. Zeng, J. Dai, W. Yao, D. Xiao, and X. Cui, Valley polarization in MoS$_2$ monolayers by optical pumping, Nature Nanotechnology \textbf{7}, 490 (2012).

\bibitem{Jones.NatNano.2013} A. M. Jones, H. Yu,	N. J. Ghimire, S. Wu, G. Aivazian, J. S. Ross, B. Zhao, J. Yan, D. G. Mandrus, D. Xiao, W. Yao, and X. Xu, Optical generation of excitonic valley coherence in monolayer WSe2, Nature Nanotechnology \textbf{8}, 634 (2013).

\bibitem{Schaibley.2016} J. R. Schaibley, H. Yu, G. Clark, P. Rivera, J. S. Ross, K. L. Seyler, W. Yao, and X. Xu, Valleytronics in 2D materials, Nature Reviews Materials \textbf{1}, 16055 (2016).

\bibitem {shafayat.valleybloch} Md. S. Hossain, M. K. Ma, K. A. Villegas-Rosales, Y. J. Chung, L. N. Pfeiffer, K. W. West, K. W. Baldwin, and M. Shayegan, Spontaneous Valley Polarization of Itinerant Electrons
Phys. Rev. Lett. \textbf{127}, 116601 (2021).

\bibitem{Shkolnikov.PRL.2002} Y. P. Shkolnikov, E. P. De Poortere, E. Tutuc, and M. Shayegan, Valley splitting of AlAs two-dimensional electrons in a perpendicular magnetic field, Phy. Rev. Lett. \textbf{89}, 226805 (2002).

\bibitem {Lai.2004} K. Lai, W. Pan, D. C. Tsui, S. Lyon, M. M\"{u}hlberger, and F. Sch\"{a}ffler, Two-Flux Composite Fermion Series of the Fractional Quantum Hall States in Strained Si, Phys. Rev. Lett. \textbf{93}, 156805 (2004).

\bibitem{Shkolnikov.PRL.2005} Y. P. Shkolnikov, S. Misra, N. C. Bishop, E. P. De Poortere, and M. Shayegan,  Observation of Quantum Hall ``Valley Skyrmions," Phys. Rev. Lett. \textbf{95}, 066809 (2005).

\bibitem{Bishop.PRL.2007} N. C. Bishop, M. Padmanabhan, K. Vakili, Y. P. Shkolnikov, E. P. De Poortere, and M. Shayegan, Valley polarization and susceptibility of composite fermions around a filling factor $\nu=3/2$, Phys. Rev. Lett. \textbf{98}, 266404 (2007).

\bibitem{Padmanabhan.PRB.2009} M. Padmanabhan, T. Gokmen, and M. Shayegan, Density dependence of valley polarization energy for composite fermions. Phys. Rev. B \textbf{80}, 035423 (2009).

\bibitem{Padmanabhan.PRB.2010} M. Padmanabhan, T. Gokmen, and M. Shayegan, Composite fermion valley polarization energies: Evidence for particle-hole asymmetry, Phys. Rev. B \textbf{81}, 113301 (2010).

\bibitem{Gokmen.Natphy.2010} T. Gokmen, M. Padmanabhan, and M. Shayegan, Transference of transport anisotropy to composite fermions, \textit{Nature Physics} \textbf{6}, 621 (2010).

\bibitem{Padmanabhan.PRL.2010} M. Padmanabhan, T. Gokmen, and M. Shayegan, Ferromagnetic Fractional Quantum Hall States in a Valley-Degenerate Two-Dimensional Electron System, Phys. Rev. Lett. \textbf{104}, 016805 (2010).

\bibitem{Gokmen2.PRB.2010} T. Gokmen and M. Shayegan, Density and strain dependence of $\nu=1$ energy gap in a valley-degenerate AlAs quantum well, Phys. Rev. B \textbf{81}, 115336 (2010).

\bibitem{Gokmen.ssc.2010} T. Gokmen, M. Padmanabhan, and M. Shayegan, Temperature dependence of piezoresistance of composite Fermions with a valley degree of freedom, Solid State Commun. \textbf{150}, 1165 (2010).

\bibitem{Young.NatPhy.2012} A. F. Young, C. R. Dean, L. Wang, H. Ren,	P. Cadden-Zimansky,	K. Watanabe, T. Taniguchi, J. Hone,	K. L. Shepard, and P. Kim, Spin and valley quantum Hall ferromagnetism in graphene, Nature Physics \textbf{8}, 550 (2012).

\bibitem {Feldman.2012} B. E. Feldman, B. Krauss J.  H. Smet and A. Yacoby, Unconventional Sequence of Fractional Quantum Hall States in Suspended Graphene, Science \textbf{337},1196 (2012).

\bibitem {Feldman.2013} B. E. Feldman, A. J. Levin, B. Krauss, D. A. Abanin, B. I. Halperin, J. H. Smet, and A. Yacoby, Fractional Quantum Hall Phase Transitions and Four-Flux States in Graphene, Phys. Rev. Lett. \textbf{111}, 076802 (2013).


\bibitem {Chung.PRM.2018} Y. J. Chung, K. A. Villegas Rosales, H. Deng, K. W. Baldwin, K. W. West, M. Shayegan, and L. N. Pfeiffer,  Multivalley two-dimensional electron system in an AlAs quantum well with mobility exceeding $2\times10^{6}$ cm$^{2}$/Vs, Phys. Rev. Materials \textbf{2}, 071001(R) (2018).

\bibitem {ShafayatAlAs.PRL.2018} Md. S. Hossain, M. K. Ma, Y. J. Chung, L. N. Pfeiffer, K. W. West, K. W. Baldwin, and M. Shayegan, Unconventional Anisotropic Even-Denominator Fractional Quantum Hall State in a System with Mass Anisotropy, Phys. Rev. Lett. \textbf{121}, 256601 (2018).

\bibitem {Jain.2007}  J. K. Jain, \textit{Composite fermions} (Cambridge University Press, Cambridge, 2007).

\bibitem{Halperin.Jain.2020} B. I. Halperin, The Half-Full Landau Level, "Fractional Quantum Hall Effects: New Developments," edited by B. I. Halperin and J. K. Jain (World Scientific Publishing Co., 2020).

\bibitem {Jain.PRL.1989} J. K. Jain, Composite-fermion approach for the fractional quantum Hall effect, Phys. Rev. Lett. \textbf{63}, 199 (1989).

\bibitem{Halperin.Lee.Read.PRB.1993} B. I. Halperin, P. A. Lee, and N. Read, Theory of the half-filled Landau level, Phys. Rev. B \textbf{47}, 7312 (1993).

\bibitem{FootenoteB} The $\Lambda$L diagram of Fig. 2(e) assumes that CFs are fully spin polarized. This is the case at the high magnetic fields where we are analyzing our data.

\bibitem{Footnote.1/2} We caution, however, that the shape of the piezoresistance in Fig. 2(d) is different from the parabolic shape seen in previous measurements \cite{Padmanabhan.PRB.2009, Padmanabhan.PRB.2010, Gokmen.ssc.2010}. Moreover, the values of $|\varepsilon|$ above which the resistance at $\nu=1/2$ saturates in Fig. 2(d) saturates is very close to where we observe full valley polarization of \textit{electrons} at $B=0$; see Fig. 1(d). (Note that there is also a rise in resistance at several FQHSs, e.g., at $\nu=2/3$, $3/5$, and $4/7$ [see Figs. 2(a-c)] at the same $|\varepsilon|\simeq 0.8 \times 10^{-4}$.) Considering the above observations, we show the data point in Fig. 2(f) at $\nu=1/2$ by an open symbol to distinguish it from data points at other fractions.

\bibitem {F1} Our self-consistent calculations of potential and charge distribution are carried out for $B=0$. However, since of interest here is the status of the 2DES in the extreme quantum limit where the electrons presumably occupy the lowest, spin-polarized LL, we assume in the calculations that the 2DES is fully spin polarized and occupies only the lowest electric subband.


\bibitem{Park.PRL.1998} K. Park and J. K. Jain, Phase Diagram of the Spin Polarization of Composite Fermions and a New Effective Mass, Phys. Rev. Lett. \textbf{80}, 4237 (1998).

\bibitem{FN} Note that, based on the $\Lambda$L diagram of Fig. 2(e), we would expect all the FQHSs to remain strong past their (last) transition into a fully valley-polarized FQHS. This is indeed experimentally observed for the sequence of FQHSs on the high-field side of $\nu=1/2$
(namely, $2/5$, $3/7$, and $4/9$) where the resistance remains small past the last transition. However, the FQHSs on the low-field side of $\nu=1/2$ (namely, $2/3$, $3/5$, and $4/7$) exhibit a rising resistance when $|\varepsilon| \gtrsim 0.8 \times 10^{-4}$. The resistance at $\nu=1/2$ [Fig. 2(d)] also shows a rise for $|\varepsilon|  \gtrsim 0.8 \times 10^{-4}$. These rises might stem from the full valley polarization of the 2D \textit{electron} system when $|\varepsilon|$ exceeds $ 0.8 \times 10^{-4}$, as implied by Fig. 1(d) data.


\bibitem{SM} See Supplemental Material for additional data and discussion (see, also, Refs.  \cite{Shayegan.APL.2003, eisenstein.prb.1990, engel.prb.1992, Leadley.PRL.1997} therein).

\bibitem{Shayegan.APL.2003}M. Shayegan, K. Karrai, Y. P. Shkolnikov, K. Vakili, E. P. De Poortere, and S. Manus, Low-temperature, in situ tunable, uniaxial stress measurements in semiconductors using a piezoelectric actuator, Appl. Phys. Lett. \textbf{83}, 5235 (2003).

\bibitem{eisenstein.prb.1990} J. P. Eisenstein, H. L. Stormer, L. N. Pfeiffer, and K. W. West, Evidence for a spin transition in the $\nu=2/3$ fractional quantum Hall effect, Phys. Rev. B \textbf{41}, 7910(R) (1990) 

\bibitem{engel.prb.1992} L. W. Engel, S. W. Hwang, T. Sajoto, D. C. Tsui, and M. Shayegan, Fractional quantum Hall effect at $\nu=2/3$ and $3/5$ in tilted magnetic fields, Phys. Rev. B \textbf{45}, 3418 (1992).


\bibitem{Leadley.PRL.1997} D. R. Leadley, R. J. Nicholas, D. K. Maude, A. N. Utjuzh, J. C. Portal, J. J. Harris, and C. T. Foxon, Fractional Quantum Hall Effect Measurements at Zero g Factor, Phys. Rev. Lett. \textbf{79}, 4246 (1997).

\bibitem {Sondhi.PRB.1993} S. L. Sondhi, A. Karlhede, S. A. Kivelson, and E. H.  Rezayi, Skyrmions and the crossover from the integer to fractional quantum Hall effect at small Zeeman energies, Phys. Rev. B \textbf{47}, 16419 (1993).

\bibitem{Liu.PRB.2014} Y. Liu, S. Hasdemir, A. W\'{o}js, J. K. Jain, L. N. Pfeiffer, K. W. West, K. W. Baldwin, and M. Shayegan, Spin polarization of composite fermions and particle-hole symmetry breaking, Phys. Rev. B \textbf{90}, 085301 (2014).

\bibitem{Zhang.PRL.2016} Y. Zhang, A. W\'{o}js, and J. K. Jain, Landau-Level Mixing and Particle-Hole Symmetry Breaking for Spin Transitions in the Fractional Quantum Hall Effect, Phys. Rev. Lett. \textbf{117,} 116803 (2016).

\bibitem{Footnote.LL.mixing} In our sample the so-called LL mixing parameter, namely the ratio of the Coulomb energy to cyclotron energy, is $\simeq 6.4$ at $\nu=1/2$. 

\bibitem {Zhu.PRB.2018} Z. Zhu, D. N. Sheng, L. Fu, and I. Sodemann, Valley Stoner instability of the composite Fermi sea, Phys. Rev. B \textbf{98}, 155104 (2018).

\bibitem {Stoner.1947} E. C. Stoner, Ferromagnetism, Rep. Prog. Phys. \textbf{11}, 43 (1947).



\end{thebibliography}

\begin{thebibliography}{99}

\bibitem{Bishop.PRL.2007} N. C. Bishop, M. Padmanabhan, K. Vakili, Y. P. Shkolnikov, E. P. De Poortere, and M. Shayegan, Valley polarization and susceptibility of composite fermions around a filling factor $\nu=3/2$, Phys. Rev. Lett. \textbf{98}, 266404 (2007).

\bibitem{Padmanabhan.PRB.2009} M. Padmanabhan, T. Gokmen, and M. Shayegan, Density dependence of valley polarization energy for composite fermions. Phys. Rev. B \textbf{80}, 035423 (2009).

\bibitem{Padmanabhan.PRB.2010} M. Padmanabhan, T. Gokmen, and M. Shayegan, Composite fermion valley polarization energies: Evidence for particle-hole asymmetry, Phys. Rev. B \textbf{81}, 113301 (2010).


\bibitem{Shayegan.APL.2003}M. Shayegan, K. Karrai, Y. P. Shkolnikov, K. Vakili, E. P. De Poortere, and S. Manus, Low-temperature, in situ tunable, uniaxial stress measurements in semiconductors using a piezoelectric actuator, Appl. Phys. Lett. \textbf{83}, 5235 (2003).

\bibitem{eisenstein.prb.1990} J. P. Eisenstein, H. L. Stormer, L. N. Pfeiffer, and K. W. West, Evidence for a spin transition in the $\nu=2/3$ fractional quantum Hall effect, Phys. Rev. B \textbf{41}, 7910(R) (1990) 

\bibitem{engel.prb.1992} L. W. Engel, S. W. Hwang, T. Sajoto, D. C. Tsui, and M. Shayegan, Fractional quantum Hall effect at $\nu=2/3$ and $3/5$ in tilted magnetic fields, Phys. Rev. B \textbf{45}, 3418 (1992).


\bibitem{Leadley.PRL.1997} D. R. Leadley, R. J. Nicholas, D. K. Maude, A. N. Utjuzh, J. C. Portal, J. J. Harris, and C. T. Foxon, Fractional Quantum Hall Effect Measurements at Zero g Factor, Phys. Rev. Lett. \textbf{79}, 4246 (1997).

\bibitem{Shkolnikov.PRL.2005} Y. P. Shkolnikov, S. Misra, N. C. Bishop, E. P. De Poortere, and M. Shayegan, Observation of Quantum Hall ``Valley Skyrmions," Phys. Rev. Lett. \textbf{95}, 066809 (2005).


\end{thebibliography}
\end{document}